# A Theory of Scientific Practice

Hisham Ghassib

ghassib@psut.edu.jo

The Princess Sumaya University for Technology (PSUT),
P.O.Box 1438, Al-Jubaiha 11941, Jordan

## Abstract

This paper avers that science is not demarcated from other disciplines by a specific unique methodology, but by its specific scientific rationality and rational grounds. In this context, the notion and structure of scientific reason are explicated. Four rational grounds of science are identified: the epistemological. Ontological, ethical and sociological grounds. They are discussed in detail within the context of classical physics, relativistic physics and quantum physics.

## 1- Introduction

Knowledge is a peculiar social product. It is the only a priori unenvisageable social product. It is so by nature. An envisageable knowledge is a contradiction in terms. Accordingly, there is no assembly-line-like method for producing knowledge. In short, there is no such thing as a specific scientific method, with specific universal steps to be universally followed. Every research leader has his own method of thinking and discovery. There is a family resemblance between the various methods, but they are quite distinct. Of course, scientists learn from each other. Usually, a novice scientist starts with his mentor's methodology, but as his research and problem-solving develops and evolves, he modifies his methodology accordingly, and this could lead him to inventing an altogether new methodology. Every methodology has its limits. When it no longer serves as a tool of discovery, it is transcended to another, generally more comprehensive, methodology. Thus, methodology is an organism that evolves. Generally speaking, we can say that every discovery and theoretical breakthrough entails a methodological breakthrough. In fact, methodology and theory are dialectically related. They are different faces of the same coin. Methodology is active theory. Each methodology is a key to an aspect of reality. We need to constantly develop methodology in order to unlock more of its aspects.

Does this mean science cannot be demarcated from other cultural practices-- from philosophical, religious, magical, astrological, mystical, alchemical, and artistic practices? Does the denial of the existence of the scientific method entail the interpenetration of various cultural constructs, and the impossibility of demarcating science from other cultural practices? (Feyerabend, 1978).

Not in the least. I aver that scientific practice is a distinct practice that can be demarcated rather precisely from other cultural practices. In that case, what are the distinguishing features of scientific practice?

## 2- A Theory of Scientific Practice

Science is not distinguished from other cultural disciplines by its methodology, but, rather, by its rationality and intellectual grounds. It entails a distinct, though evolving, rationality, and is based on a set of specific grounds. It entails the existence of a scientific reason, and a set of grounds-- epistemological, ontological, ethical and sociological grounds.

## 2.1- The Epistemological Grounds of Scientific Practice

a- Science is a self-contained whole-- a closed epistemological universe. It does not recognize any epistemological reference point beyond itself. It does not recognize any authority outside itself. Its criteria of validation and truth lie within itself. Its very truth lies within itself. Its only reference point is scientific reasoning. The essence of scientific reasoning is the necessary dialectical synthesis between mathematized theorization and precise experimentation and measurement (Ghassib, 1992). In this regard, we are justified in talking about scientific reason, a specific reason with a specific structure. We shall later on devote a whole section for a detailed treatment of scientific reason. The essence, spirit and only reference point of science is scientific reason. It is what gives science its unique character. It informs the infinite variety of methods, which we call scientific methods. Thus, the unifying element of science is not the scientific method, which is actually a myth, but, rather, scientific rationality, scientific reason.

b- Scientific knowledge is theoretically structured. Theory is the necessary form of scientific knowledge. Thus, the latter is necessarily structured and ordered. It is necessarily systematized and structurally constituted. Its ideal image is mathematics-- mathematical constructions. Thus, it is necessarily axiomatized. Its basic form is mathematico-conceptual constructions on the basis of a set of axioms. However, unlike mathematical axioms, which are ideal and arbitrary, and unlike religious axioms, which are absolute and absolutely certain, scientific axioms are relative and real. They are relative in the sense that they are conditioned and limited. They are true under certain conditions and within specific limits. They are real in the sense that they are constantly subject to critique, empirical verification and revision. They are not static, but constantly evolve.

c- In science, there is always appearance and reality. Both classical physics and quantum physics acknowledge that, even though the latter has transformed the way we view the relationship between them. Appearance is principally subjective, even though it appears objective. It gives the illusion of being objective. In so far as it is considered objective, it is highly deceptive. Science views it as a mere token of reality. It is a mere sign that points to an underlying reality. It views it as the result of the action of the underlying reality onto the senses. Even in quantum mechanics, the underlying reality is considered qualitatively more objective than the appearance, even though it is not considered wholly objective (Heisenberg, 1949). In science, the senses are considered approximate and defective measuring instruments. Thus, they are not adequate for exploring reality. We need to construct more accurate measuring devices on the basis of mathematized theory. That is, we need scientific reason to discover and probe reality. Scientific reason is our main instrument to probe reality. It is our eye onto reality. We do not see reality with our eyes, but with our scientific reason.

d- Scientific vision is in many respects Platonic. It deals with ideal representations and abstractions. The basic unit of scientific vision is the possible ideal physical system. This is more of a model of reality than a real representation. It is abstracted from reality. In this sense, it bears resemblance to reality, but it is basically an ideal representation of reality. It is usually approximately realizable in reality via experiment. It is described in terms of certain properties and constructs. These are specifiable in terms of mathematics and measurement. Their essential form of representation is mathematics and entails precise measurement. The laws of nature are durable relations between these primary qualities and constructs. Science is basically a quest for new such primary qualities and constructs (physical quantities) and the relations between them, and for their manifestations in specific instances. Thus, the laws of nature are also ideal. They are not merely relations governing the real world, but are universal active forms governing all possible ideal physical systems-- that is, the space of all possible ideal situations. Thus, in a sense, the laws of nature are prior to nature and above it. They are more akin to universal Platonic forms, which are not only more real than nature itself, but the source of its reality (Ghassib, 2010). Experiment is the necessary bridge between the general ideal world and the real world (Bhaskar, 1978).

e- Scientific practice is intrinsically critical. It does not accept events, phenomena and things at their face value. It does not readily accept the given. It constantly questions what is given. Crude direct empiricism is alien to it. It critiques the given to reveal its reality. It never treats a given item in isolation from its conditions of existence. It constantly questions, critiques, and looks for connections and grounds. In this respect, scientific practice is intrinsically causal.

f- Scientific practice is intrinsically holistic. Context is all important to it. It seeks deeper unities. It realizes that wholes manifest themselves in seemingly isolated fragments. Thus, it is in a constant quest for identity connections, contexts, conditions and identities (Bohm, 1985).

## 2.1.1 Scientific Reason

The essence of scientific reason, which is the cornerstone of scientific practice, is the dialectical relationship between mathematized theorization and precise experimentation and measurement. However, this relationship is a complex multi-faceted relationship. It consists of a number of organically inter-related methodological operations, that constantly transform each side as such, in addition to transforming one side into the other. The main methodological operations constituting this relationship are the following:

2.1.1.1- *Induction:* This is supposed to be the very essence of modern scientific practice. But, in fact, it is one of many fundamental methodological operations, and obtains its validity from the whole. It is the operation of legitimate generalization. That is, it is the movement from the particular to the general-- from a limited and finite set of measurements to a general statement. It is normally not a routine, straight-forward, mechanical operation, but, rather, a very complex, imaginative and creative process. In addition to employing various, sometimes sophisticated, mathematical techniques, this operation sometimes entails building new models, and is often guided by new physical principles. In fact, induction is a form of reading, or interpretation. It is an interpretation of a set of measurements in terms of possible real situations, using mathematical, physical and philosophic models and principles. In the history of science, induction is a family of operations, rather than one specific and unique operation. This is clearly epitomized in such well-known examples as: Ohm's law, Kepler's laws of planetary motions (Stephenson, 1987), and Rutherford's discovery of the Atomic nucleus (Kragh, 1999).

2.1.1.2- *Deduction:* This is supposed to be the opposite of induction-- that is, the movement from the general to the specific or particular. In a sense, it is the reduction of the general universal form of a law of nature to a specific manifestation of it. It is the concretization of theory. Deduction is the logico-mathematical development of a theory, the purpose of which is to make the theory functional. For the theory to become comprehensive and to appropriate reality, it must be developed in this way. Thus, deduction is the logico-mathematical mechanism, whereby theory appropriates the concrete-- reality. This is clearly illustrated by such renouned instances of deduction as Galileo's discovery of the parabolic nature of parabolic paths (Galileo, 1954), Newton's deduction of Kepler's laws of planetary motions (Bernard Cohen,1983), Hamilton's deduction of classical mechanics (Hankins, 1985), Maxwell's deduction of the existence of electromagnetic waves (Ghassib, 1988), Einstein's deduction of the Lorentz transformations from the basic axioms of special relativity (Einstein, 1998), the deduction of the existence of black holes from general relativity (Hawking et. Al., 1987), Scrodinger's deduction of the spectrum of the hydrogen atom from his wave equation (Ludwig, 1968), and Dirac's deduction of the existence of anti-matter from his wave equation (Kvasnica, 1964), amongst others.

*The Conditions and Functions of Deduction*:

Deduction, which, in modern physics, flows from the mathematization of nature, and expressing the laws of nature in terms of differential equations, has endowed modern physics with an uncanny, almost magical, power. It is an incredible undreamt of power. It has turned scientific theories into exceedingly powerful engines of knowledge production and unification. Deduction lies at the basis of explanation and prediction (Balashov et. al., 2002). A phenomenon or event is explained when it is deduced mathematically and logically from a set of principles. Thus, it could be said that Newton's deduction of Kepler's laws of planetary motions, and Schrodinger's deduction of the spectrum of the hydrogen atom from his wave equation were explanations of these phenomena. When Hans Bethe deduced the enormous emergy output of the sun from nuclear reactions, he furnished an explanation of the existence of the sun for the first time in history (Kragh, 1999). The blueness of the sky was explained the moment wave theorists were able to deduce the law of scattering of light waves from air molecules (Sommerfeld, 1964). However, deduction also lies at the basis of prediction. The glory of modern physics flows out of its tremendous predictive power. When new unknown results are deduced and properly interpreted, they are called predictions. Thus, the deductions of time dilation, length contraction, black holes, the expansion of global space, anti-matter and suchlike are predictions. Theories are supposed to be explanatory and predictive, if they are to be taken seriously. These functions are essential features of a successful theory. Thus, despite its formidable equations and concepts, Einstein's general relativity was quickly accepted because it could satisfactorily explain-- that is, mathematically deduce-- the precession of the perihelion of mercury, and because it predicted the precise impact of gravitation on light beams (Eisenstaedt, 2006). Thus, deduction is the very spirit of theory.

With the calculus, deduction has become truly a substantial development of theory, a positive process, and not a mere logical or syllogistic exercise. However, for a theory to be deductive, it must be endowed with a sufficiently high degree of logical coherence.

Otherwise, deduction, as well as explanation and prediction, lose their meaning. For deduction to give specific, unambiguous results that could be tested, a theory must be logically coherent to a large extent. It may be that we can never construct a perfectly logically coherent theory. Thus, it is a matter of degree. The limits of logical coherence are also limits on the efficacy of the theory. Thus, for example, the Big Bang theory of the Universe breaks down at the singularity-- the so-called moment of creation (Hawking, 1998). That is, it loses its logical coherence at this point. This sets a limit on the efficacy of the theory and motivates a search for a more comprehensive theoretical framework, that would cope with the singularity or eliminate it altogether. Accordingly, the problem of logical coherence has become a major problem in theoretical physics. This has reached its zenith in quantum field theory (Ghassib, 2009). One of the motivations for developing string theory has been to overcome the problem of logical coherence in quantum field theory (Green, 1999). This poses the question: In view of this problem, is there a limit to universality and comprehensiveness of a theory? Is a theory of everything impossible? If so, is it a mere epistemological limit, related to the limits of human brain power, or is it an ontological limit, related to the nature of the Universe? (Weinberg, 1992).

*2.1.1.3- Tests*:

By testing a theory, we mean developing it by deduction-- that is, by applying it to a possible reproducible ideal system-- reproducing the system in reality (usually in the lab), and comparing the theoretical deductive results with the experimental results. Thus, it involves deduction, prediction and experimentation. Theories always seek to be tested. Tests are necessary not only for validation, but also for specifying the theory and its path of development. Tests have acquired a crucial importance in the 20th century, in view of the barrage of unfamiliar, almost crazy, ideas that century witnessed, such as time dilation (Rosser, 1971), and De Broglie's matter waves (Eisberg, 1961).

2.1.1.4- *Hypothesis:*

Often, the prevalent theoretical framework is not sufficiently specific to lend itself to direct explanatory deduction. It usually provides a space of possibilities, rather than specific and definite deductions. Sometimes, it proves to be inadequate and incomplete. In those cases, explanatory models are constructed out of a combination of existing theoretical elements and reasonable assumptions, which need to be justified and confirmed later. Such models are called hypotheses. When logically and experimentally confirmed, these hypotheses are incorporated into the theoretical corpus. Thus, they are a means of enriching and expanding the theoretical edifice, and even concretizing it. However, during periods of crises and revolutions, they could be means of undermining and demolishing the existing theoretical edifice. For example, in 1894, Max Planck turned his attention to the so-called problem of the black body radiation. He applied the grand classical theoretical corpus to the radiation trapped in a cavity at equilibrium, and analyzed it in terms of an interaction between the trapped radiation and presumed field oscillators constituting the walls of the cavity. To start with, he made it a point to stick to those parts of the classical corpus, which he deemed reliable, such as classical mechanics, thermodynamics and classical electrodynamics. However, he soon realized that this theoretical core was inadequate for explaining the spectrum of black body radiation. So, he realized he had to introduce Boltzmann's statistical mechanics, even though he abhorred it. This led him to the brink of a successful explanation, at the turn of the 20th century, but not quite. He realized that he had to introduce a somewhat alien hypothesis-- the so-called quantum hypothesis. The latter stipulates that energy is absorbed and emitted by the field oscillators discretely or discontinuously, in packets, the energy of each of which is proportional to the frequency of the absorbed or emitted radiation. Thus, he had to introduce an element from the outside. This element was not only an outside element, but proved to be contradictory to some basic principles of the classical theoretical corpus (Kuhn, 1978). In fact, by 1925, it led to undermining this grand corpus, replacing it with quantum mechanics (Hoffmann, 1986). Another pertinent example is Einstein's axiom of the constancy of the speed of light. This started out as a crazy hypothesis, but its incredible successes in explaining and predicting phenomena soon turned it into a fundamental principle of physics.

2.1.1.5- *Gedanken Experiment:*

The Platonic feature of scientific practice reveals itself most conspicuously in gedanken or thought experiments, which have played a crucial role in the history of scientific innovation. A gedanken experiment is an ideal physical system, which is most often unrealizable in practice under the prevailing conditions and circumstances . It has various functions, such as clarifying a principle by concretizing it, critiquing a concept or principle, clarifying its actual meaning, showing its reasonableness, absurdity or contradictions and problems, and revealing its nature. Well-known examples are Galileo's moving ship experiment ( Barbour, 2001), Newton's cannon experiment (Bernard Cohen, 1985), Maxwell's demon (Kuhn, 1978), Einstein chasing a light beam (Isaacson, 2007), the twin paradox (French, 1968), Heisenberg's microscope (Bohm, 1979), Schrodinger's cat, Wigner's friend and the EPR experiment (Gribbin, 1995).

2.1.1.6- *Critique:*

Critique plays a very important and durable part in scientific practice. What do we mean by critique? Critique entails revealing the hidden structure, internal and external contradictions and relations, limits, grounds, lineage and potentialities of an idea. It often entails deconstructing and reconstructing an idea, pointing out structural defects and absences in the idea critiqued, and evaluating and assessing various aspects of an idea. Amongst the most crucial instances of critique in the history of science are Aristotle's critique of Plato, his contemporaries and predecessors (Aristotle, 1978), Al-Hazen's critique of Greek optics (Rashed, 2003), Galileo's critique of Aristotle (Drake, 1999), Newton's critique of Descartes (Koyre, 1957), Mach's critique of Newtonian mechanics (Smart, 1964), Einstein's critique of classical kinematics and Bohr's and Heisenberg's critique of calssical concepts (Heisenberg, 1971; Beller, 1999).

2.1.1.7- *Dialectical Synthesis*:

This is an often neglected methodological operation. But, it is indispensable. Its crucial importance is revealed when we pose the following question: How are theories constructed? I aver that theory construction cannot be comprehended without this operation. Induction and deduction are not adequate to explain this major part of scientific practice. By dialectical relation, I mean a necessary, existential, transformational and contradictory relation. Dialectical synthesis is merging different, contradictory elements into new, more encompassing elements, resolving the contradictions in the process. Theory develops principally in this manner. It develops by resolving contradictions via dialectical synthesis. Theories do not emerge in isolation from each other, but in deep connection. Theories give birth to each other; they generate each other dialectically. Thus, theory is an evolving organism (Ghassib, 1988). The major instances of dialectical synthesis are Galileo's synthesis of Archimedes' mathematical method with Platonic idealism, Neton's grand synthesis of Galileo's terrestrial mechanics and Kepler's celestial mechanics (Bernard Cohen et. al., 1983), Hamilton's grand synthesis of classical mechanics (Lanczos, 1970), Maxwell's Unification of Electricity and Magnetism (Ghassib, 1988), Boltzmann's statistical mechanics (Kuhn, 1978), Einstein's special relativity (Jammer, 2006), Minkowski's spacetime (Lawden, 1968), Einstein's general relativity (Pais, 2005), the Schrodinger equation (Ghassib, 1983), Dirac's relativistic wave equation (Kvasnica, 1964), and the electroweak synthesis (Pagels, 1986).

2.1.1.8- *Significant Observation*:

By significant observation, we mean reading a grand fact or principle in a simple small quantitative difference. Like any ability to read, this requires a highly trained and qualified mind-- a fully prepared, theoretically structured mind. It is more of a discursive than a raw intuitive power. Significant observation reveals itself so clearly in such celebrated examples as Eratosthenes' measurement of the circumference of the Earth (Sarton, 1987), Newton's analysis of circular motion (Bernard Cohen, 1985), Maxwell's unification scheme, and Einstein's equivalence principle.

## 2.2- The Ontological Grounds of Scientific Practice

I want to explore the crisis of meaning in physics by making a detailed fundamental comparison between classical physics and quantum mechanics, ontologically, methodologically, and logically.

I will start by addressing the problem of objective reality in physics.

This can be approached via the question of ontology in physics. By ontology, we mean the theory of being.

It seems to me that scientific practice, at least within the context of classical physics, entails the following ontology:

1- There is a mind-independent reality underlying all phenomena and events.

2- This reality is knowable via scientific rationality.

3- Ultimate reality is ordered-- that is, it is structured and governed by natural laws. We must not take the meaning of natural laws for granted. We will explore their meaning later when we talk about quantum mechanics.

4- Nature is an infinite physical system consisting of interacting material components. Whether these components exist on their own or in relation to each other and to the whole is an open question. The laws of nature are obeyed by all subsystems. It is assumed that they are obeyed by the whole--the Universe-- as well.

5- The laws of nature are infinitely universal, in the sense that they apply universally to all possible, including ideal limiting, physical systems. These laws condition nature, but are not conditioned by it.

6- Interactions between material components are the ultimate cause of events and changes in the universe. Interaction is the essence of scientific explanation.

7- There is a contradictory dislocation between appearance and reality. What bridges the gap is the theory of interaction. In this respect, interaction is truly the essence of modern physics.

This analysis should be concretized by inquiring about how possible physical systems are characterized. In classical mechanics, they are generally characterized by primary qualities, or physical quantities. These are objective primary properties, whose representational form is mathematics, and which are measurable without theoretical limits. The objective of science is to discover these primary properties, discover the connections between them, and specify their values for specific systems using these relations, which we call the laws of nature.

Does this ontology accord with quantum mechanics?

Quantum mechanics does not start with well-defined physical quantities. Rather, it starts with state vectors, which are vectors in Hilbert space. Then, it introduces linear Hermitian operators, which represent so-called observables, rather than physical quantities as such. These operators act on state vectors to yield a spectrum of eigenvalues, which are possible values of the observables. The act of measurement selects a specific value. In general, the state vector is a linear combination of the eigenfunctions, with the coefficients representing the probability amplitudes of the various eigenvalues and eigenfunctions. The act of measurement makes the state vector collapse into a specific eigenfunction. Not all observables are compatible with each other, which means that they cannot all be specified simultaneously. This incompatibility depends on the experimental conditions (Bohm, 1979).

Notice that this scheme accords right from the start a special status to the act of measurement. What are the implications of this formulation?

1- Reality is not purely objective. It is measurement dependent. Is measurement a subjective act? Is it basically an interaction with a macroscopic object? Or, is it theory laden, and, therefore, an embodiment of human reason? If the former, then some sort of holism is implied. If the latter, then some sort of subjective idealism is implied (Heisenberg, 1979; Reichenbach, 1975).

2- Since human acts play a role in constituting reality, to what extent can we say that reality is knowable?

3- Since probability plays such an important role, to what extent can we say that reality is ordered? The laws of nature are no longer causal relations between objective properties, but, rather, relations between observables, which cannot be defined without human acts. If so, to what extent can we say reality is ordered?

4- The concept of possible physical system is retained. However, the relationship between the sub-system and the whole becomes much more problematic (Bohm et. al., 1993).

5- Physical laws retain their universal character, but lose their self-sufficiency. They do not stand on their own, but require random subjective lawless acts for their efficacy.

6- Interaction retains its causal efficacy, but does not govern all events and changes. Acausal randomness plays a role (Bohm, 1957).

7- In quantum mechanics, there is a marked dislocation between appearance and reality. However, reality no longer completely underlies appearance. The latter underlies the former as much as the former underlies the latter. Reality is shaped by appearance as much as appearance is shaped by reality. The classical asymmetry disappears in quantum mechanics.

To start with, De Broglie and Schrodinger wanted to explain atomic phenomena in terms of a classical field equation, without violating classical ontology. But, they were not given the chance to develop their project (Beller, 1999). In particular, the latter developed the double solution interpretation of wave mechanics, but was terrorized into abandoning it for 25 years (Hiley et. al., 1997). Einstein was not prepared to renounce classical ontology either. He tried to show that quantum mechanics was an incomplete description of nature, and that it was a statistical theory. Thus, there was a deeper theory, which accorded with classical ontology, and which underlay quantum mechanics. He offered a real challenge to the prevalent formulation, but was vehemently opposed by the Copenhagen group (Lindley, 2007). Instead of scrutinizing his objections with an open mind, they employed all sorts of tricks to confute him, lightly dismiss him and isolate him. No doubt, quantum mechanics posed many challenges to classical ontology. However, instead of trying to overcome them, they were eager and ready to throw the baby with the tub water. It was pure ideology and dogma. Obviously, extra scientific factors were at work here. They were not ready to acknowledge the philosophical arbitrariness that underlay their formulation, but considered it the only scientifically possible formulation. It was science put in the service of ideology. In 1952, David Bohm offered a causal realistic formulation of wave mechanics, which seemed to be as good as the orthodox formulation. It was similar to De Broglie's formulation. Even though it has endured, it is still marginal. Likewise with all the formulations that were proposed later. The orthodox formulation is still prevalent, despite the challenges. The desire to renounce classical realism is still there. But, can we afford to persist in renouncing it? What are the consequences of such a renunciation  on the whole enterprise of physics? When the physicist formulates a theory or an explanation, or when he performs an experiment, or when he makes a prediction, he seems to tacitly assume classical realism. Thus, does scientific practice contradict quantum ontology? If there is such an irreconcilable contradiction, that could spell disaster for the very enterprise of physics. It is tantamount to self-destruction or suicide. Also, can we afford to lose the objectivity of the laws of nature? In the orthodox formulation, the laws of nature are no longer objective relations between objective physicals quantities, but, rather, relations between acts of measurement. That leads to the renunciation of the very concept of reality. Bohr seems to acknowledge this consequence, but he seems to revel in it (Bohr, 1998). Heisenberg seems to consider reality a mere sea of potentialities (Heisenberg, 1971). It needs human acts to transform it into actuality. Once again, we have human act constituting reality. Back to good old Kant? But, they do not acknowledge that. Sartre could afford to propose such a dualism (Sartre, 1972). But, can science afford to? What are the consequences of such a view? I think the problem has not been addressed with the rigor it deserves.

The orthodox quantum physicists start with measurement, rather than with objective reality, and end up with objectifying measurement, and subjectifying reality. They start with measurement, and end up with measurement. This indicates their adherence to some amalgam of positivism and subjective idealism. On the other hand, David Bohm started with well-tested theory, with Schrodinger's equation, and ended up with a causal realistic formulation. It seems that the starting and end-points depend on the philosophy you start with. This needs further exploration.

Methodologically, quantum mechanics seems to turn classical physics on its head.

Classical mechanics starts with definite physical meaning, gives it mathematical expression and arrives at general mathematical equations as a culmination of physical meaning. Quantum mechanics starts with general mathematical equations, and then starts looking for physical meaning, which it never seems to find. It is a never-ending quest for physical meaning. Even the relationship between law and phenomenon has been reversed. In classical physics, laws explain phenomena. In quantum mechanics, in a sense, phenomena explain laws. Phenomena are not there to be explained, but they are used to illuminate the meaning of laws. It is a topsy turvy world .

Another methodological point is that, whereas in classical mechanics, abstraction flows out of the sensed and experienced world, and therefore never loses touch with the sensible in terms of meaning and compliance, in quantum mechanics, it is divorced from the sensed world. But, at least, it establishes some sort of connection with it via elaborate experiment. Unfortunately, things have deteriorated in the last thirty years. For, even the experimental connection has been severed in string theory, loop quantum gravity, and related areas (Smolin, 2006).

A third methodological point is that, in 20th century and 21st century physics, mathematical principles have become hegemonic, and have come to replace physical principles. This transformation started with Einstein, and first became manifest in his scientific career. To start with, Einstein was averse to complex mathematization. He insisted on simple physical principles with the minimum of mathematics. That is why he was not enthusiastic about Hermann Minkowski's development of special relativity. However, he soon realized that he could not make any progress towards a relativistic theory of gravitation without resorting to sophisticated mathematics. He succeeded in combining both views in constructing his theory of general relativity, which he achieved by combining simple physical principles with sophisticated mathematical principles. After that, he became so enamored and fascinated by sophisticated mathematics that he attempted to construct his unified field theory purely with very abstract geometrical principles. This was soon to become a major trend in all of 20th century physics. This has become particularly prominent in the new theoretical programs, such as m-theory, loop quantum gravity and other quantum gravitational formulations (Callender et.al., 2001). The symmetry principles of these schemes, including the holographic principle, can hardly be called physical principles. They are more accurately described as generalized geometrical principles. This development raises the question of meaning in recent physics.

Recent physics has also highlighted the problem of coherence in physical theories. As theories become more mathematical and more sophisticated, it becomes increasingly difficult to construct a stable, logically coherent, theory. This problem arose particularly forcefully in quantum field theory. One of the motivations for moving towards superstring theory was precisely to overcome the logical inconsistencies of quantum field theory. The latter may have solved some of these inconsistencies, but at the expense of losing experimental touch with reality. The problem is how to construct a logically coherent theory with full explanatory and predictive power. This is the problem facing theoretical physics today. Is this problem ontological? That is, is it rooted in reality itself? Is there an element of irrationality in reality itself? Or, is it an epistemological problem, rooted in the limitations of human reason?

## 2.3- **The Ethical Grounds of Scientific Practice**
Scientific practice is not value free. It is ethically structured. It presupposes a number of scientific ethical norms. Without these ethical principles, it will collapse. Foremost among these scientific ethical principles are: honesty in conducting experiments and constructing theories, integrity, precision, meticulousness, thoroughness, not jumping into conclusions, acknowledging past and contemporary contributions, and exhibiting team spirit. These are ethical conditions of proper scientific practice.

## 2.4- **The Sociological Grounds of Scientific Practice**
In this section, I do not intend to address the problem of the socio-historical conditions for the existence and persistence of science and for turning it from a marginal into a major and central enterprise. I have developed a comprehensive theory of such conditions elsewhere (Ghassib, 1993). Here, I wish o outline the socio-educational grounds of scientific practice. In a previous work, I have constructed a theory of knowledge production. In this theory, I have introduced the important concept of epistemic heritage, which is the raw material of the process of knowledge production. A scientist produces knowledge by working on, and with, his epistemic heritage. The epistemic heritage is generally a complex scheme and structure of concepts, implements, operations and practices. Conceptually, it is never a perfectly logically coherent system, but always possesses a degree of incoherence and inhomogeneity. Its ideal is a theory of everything, or a logical system akin in coherence to Euclidean geometry. However, in actuality, it is ridden with a degree of logical incoherence, a degree of logical inhomogeneity, contradictions, defects, gaps, uncertainties, ungrounded assumptions, conceptual haziness and unrealized potentialities. The scientist works on this complex edifice and with its elements, and systematically interacts with the object of knowledge with his epistemic heritage, in order to solve its contradictions, reduce its logical incoherence and inhomogeneity, fill in the gaps, remove the defects, clarify the concepts, ground its propositions, realize its potentialities and develop its themes. To be able to do that, a scientist must satisfy the following socio-educational requirements and conditions:

1- He must be qualified to deal with his epistemic heritage. That is, he must have gone through a thorough process of delving deeply into his epistemic heritage, understanding it and knowing its structure and problems.

2- He must have acquired the necessary mathematical, linguistic, conceptual and technical skills to work with it and on it.

3- He must have acquired the skill to get access to the sources of new information and knowledge, to update his knowledge, and to keep track of the latest developments in his field.

4- He must be emotionally committed to his scientific work, and be self-motivated to do the necessary hard work required for knowledge production. Science is a mission, rather than a routine career.

5- He must be trained to adhere to scientific values and norms of conduct.

6- He must be an active member of a scientific community, sharing with it an institutional framework, a common language, a communicative network, a common epistemic heritage, and common problems and concerns.

7- His scientific work must be conducted in the context of a research program, to which he is emotionally and intellectually committed. He could be either the leader and initiator of the research program or an active practitioner. Research programs are not static, of course, but evolve.

In short, for a scientist to be a knowledge producer, he must be qualified and prepared intellectually, practically, emotionally and morally.

## 3- Conclusion

Scientific practice exhibits an infinite variety of methods and techniques. It is a creative activity, that constantly transcends constraints and limits. However, it is grounded in epistemological, ontological, ethical and socio-educational grounds, all of its own. It is demarcated from other human practices not by a fixed unique method, but by a specific scientific rationality or scientific reason, a specific set of epistemological principles, a specific family of ontologies, a specific ethic and specific socio-educational types.

## 4- References: